# Minimization Strategies for Maximally Parallel Multiset Rewriting Systems

Artiom Alhazov[a], Sergey Verlan[b,a,]

[a]*Institute of Mathematics and Computer Science*
*Academy of Sciences of Moldova, Academiei, 5, MD-2028, Moldova*
[b] *LACL, Département Informatique, Université Paris Est,*
*61 av. Général de Gaulle, 94010 Créteil, France*

**Abstract**

Maximally parallel multiset rewriting systems (MPMRS) give a convenient way to express relations between unstructured objects. The functioning of various computational devices may be expressed in terms of MPMRS (e.g., register machines and many variants of P systems). In particular, this means that MPMRS are computationally complete; however, a direct translation leads to quite a big number of rules. Like for other classes of computationally complete devices, there is a challenge to find a universal system having the smallest number of rules. In this article we present different rule minimization strategies for MPMRS based on encodings and structural transformations. We apply these strategies to the translation of a small universal register machine (Korec, 1996) and we show that there exists a universal MPMRS with 23 rules. Since MPMRS are identical to a restricted variant of P systems with antiport rules, the results we obtained improve previously known results on the number of rules for those systems.

*Keywords:* Multiset rewriting, Universal computations, Small register machines, P systems, Symport, Antiport

## 1. Introduction

Multiset rewriting presents a convenient way to express chemical reactions. Indeed, there is a direct correspondence between chemicals and multisets, as well as between reactions and multiset rewriting. Some additional properties of the reactions' environment might be expressed by an additional control over the rewriting. This idea was heavily exploited and different multiset rewriting systems were proposed, we only mention here the Abstract Rewriting Systems on Multisets [16], the Chemical Abstract Machine [4] and the Gamma language, first considered in [2] (see also a survey in [3]).

One of natural controls that can be added to the multiset rewriting is maximal parallelism. This roughly corresponds to the idea of waiting until the chemical system reaches a stable state, *i.e.*, no more rules can be applied, for a particular step. More precisely, during a rewriting step of a maximally parallel multiset rewriting system (MPMRS) all rules that can be applied together should be applied.

When MPMRS are used for computational purposes, it is often useful to separate the data, which is transformed, from the program that does not change, *i.e.*, to permit to have an unbounded number of occurrences of symbols corresponding to the data and bound the number of occurrences of symbols

[*]Corresponding author
*Email addresses:* `artiom@math.md` (Artiom Alhazov), `verlan@univ-paris12.fr` (Sergey Verlan)

corresponding to the program. In this case the program might be represented as a set of states and transitions (that modify the data) when moving between them. We shall call MPMRS having the above property finite state maximally parallel multiset rewriting systems (FsMPMRS).

MPMRS systems serve as a basis for P systems that were introduced by Gh. Păun in [11] as distributed parallel computing devices of biochemical inspiration. These systems are inspired from the structure and the functioning of a living cell. The cell is considered as a set of compartments (regions) nested one in another and which contain objects and evolution rules. Membranes are separators of regions; they may serve as communication channels between the regions. The basic framework specifies neither the nature of these objects, nor the nature of rules.

Numerous variants specify these two parameters by obtaining many different models of computing, see [17] for a comprehensive bibliography. One of these variants, P systems with *symport/antiport*, was introduced in [10]. This variant uses one of the most important properties of P systems: the communication. This property is so powerful, that it suffices by itself to reach the computational power of Turing machines only by moving objects between the regions. These systems have two types of rules: symport rules, when several objects go together from one region to another one, and antiport rules, when several objects from two regions are exchanged. In spite of a simple definition, they are able to compute all Turing computable sets of numbers, [10]. Several subsequent works have been dedicated to improve this result with respect to both the number of membranes used and the size of symport/antiport rules used across the membranes. We refer to [14] for a survey of these investigations.

Since symport/antiport systems compute all recursively enumerable sets of numbers, it is possible to construct a universal symport/antiport P system, *i.e.*, a fixed system that will compute any partially recursive function if a corresponding input is provided. The article [5] constructs such a system having only 30 antiport rules. This result is based on a result from [8] where a universal register machine with 32 instructions is constructed.

Antiport P systems with one membrane and where the environment contains all objects (as considered in [5]) correspond in a direct manner to MPMRS. In fact, any exchange rule $(u, out; v, in)$ of an antiport system becomes a multiset rewriting rule $u \to v$ and in both cases the application of rules is maximally parallel. Moreover, most of the examples from the literature, in particular computationally complete systems, are FsMPMRS.

In this article we show that there is a universal (Fs)MPMRS with 23 rules. Thus we improve the result from [5] and we obtain a universal antiport system with the same number of rules. This result is quite surprising, because the machine from [8] that was the starting point of our construction has 25 computational branches. We also introduce a graphical notation for FsMPMRS that shows the functioning of the system and present different rule minimization strategies for (Fs)MPMRS based on encodings and structural transformations. We also continue the discussion of the relation between the number of rules and their size started in [5].

This article is an improved version of [1].

## 2. Definitions

We recall here some basic notions of formal language theory we need in the rest of the paper. We refer to [15, 12, 13] for further details.

We denote by $\mathbb{N}$ the set of all non-negative integers. Let $O = \{a_1, \ldots, a_k\}$ be an alphabet. A *finite multiset* $M$ over $O$ is a mapping $M : O \longrightarrow \mathbb{N}$, *i.e.*, for each $a \in O$, $M(a)$ specifies the number of occurrences of $a$ in $M$. The size of the multiset $M$ is $|M| = \sum_{a \in O} M(a)$. A multiset $M$ over $O$ can also be represented by any string that contains exactly $M(a_i)$ symbols $a_i$ for all $1 \leq i \leq k$, e.g., by $a_1^{M(a_1)} \ldots a_k^{M(a_k)}$, or else by the set $\{a_i^{M(a_i)} \mid 1 \leq i \leq k, M(a_i) > 0\}$. For example, the multiset over $\{a, b, c\}$ defined by the mapping $a \to 3, b \to 1, c \to 0$ can be specified by $a^3 b$ or $\{a^3, b\}$. An empty multiset is represented by $\lambda$.

We may also consider mappings $M$ of form $M : O \longrightarrow \mathbb{N} \cup \{\infty\}$, *i.e.*, elements of $M$ may have an infinite multiplicity; we shall call them *infinite multisets*.

Let $x$ and $y$ be two multisets over $O$. Then $x$ is called a submultiset of $y$, written $x \leq y$ or $x \subseteq y$, if and only if $x(a) \leq y(a)$ for all $a \in O$. The sum of $x$ and $y$, denoted by $x + y$ or $x \cup y$, is a multiset $z$ such that $z(a) = x(a) + y(a)$ for all $a \in O$. The difference of two multisets $x$ and $y$, denoted by $x - y$, provided that $y \subseteq x$, is the multiset $z$ with $z(a) = x(a) - y(a)$ for all $a \in O$. A projection of a multiset $M$ over a set $O$ is denoted by $\pi_O(M)$.

*2.1. Register machines*

A deterministic *register machine* is the following construction [8, 9]:

$$M = (Q, R, q_0, q_f, P),$$

where $Q$ is a set of states, $n$ is the number of registers, $q_0 \in Q$ is the initial state, $q_f \in Q$ is the final state and $P$ is a set of instructions (called also rules) of the following form:

1. $(p, [RiP], q) \in P$, $p, q \in Q, p \neq q, 0 \leq i \leq n-1$ (being in state $p$, increase register $i$ and go to state $q$).
2. $(p, [RiM], q) \in P$, $p, q \in Q, p \neq q, 0 \leq i \leq n-1$ (being in state $p$, decrease register $i$ and go to state $q$).
3. $(p, \langle Ri \rangle, q, s) \in P$, $p, q, s \in Q, 0 \leq i \leq n-1$ (being in state $p$, go to $q$ if register $i$ is not zero or to $s$ otherwise).
4. $(p, \langle RiZM \rangle, q, s) \in P$, $p, q, s \in Q, 0 \leq i \leq n-1$ (being in state $p$, decrease register $i$ and go to $q$ if successful or to $s$ otherwise).
5. $(q_f, STOP)$ (may be associated only to the final state $q_f$).

We note that for each state $p$ there is only one instruction of the types above.

A configuration of a register machine is given by the $(k+1)$-tuple $(q, n_1, \ldots, n_k)$, where $q \in Q$ and $n_i \in \mathbb{N}, 1 \leq i \leq k$, describing the current state of the machine as well as the contents of all registers. A transition of the register machine consists in updating/checking the value of a register according to an instruction of one of types above and by changing the current state to another one. We say that the machine stops if it reaches the state $q_f$. We say that $M$ computes a value $y \in \mathbb{N}$ on the *input* $x_1, \ldots, x_n$, $x_i \in \mathbb{N}, 1 \leq i \leq n \leq k$, if, starting from the initial configuration $(q_0, x_1, \ldots, x_n, 0, \ldots, 0)$, it reaches the final configuration $(q_f, y, 0, \ldots, 0)$.

It is well-known that register machines compute all partial recursive functions and only them [9]. For every $n \in \mathbb{N}$, with every register machine $M$ having at least $n$ registers, an $n$-ary partial recursive function $\Phi_M^n$ (computed by $M$) is associated. Let $\Phi_0, \Phi_1, \Phi_2, \ldots$, be a fixed admissible enumeration of the set of unary partial recursive functions. Then, a register machine $M$ is said to be *strongly universal* if there exists a recursive function $g$ such that $\Phi_x(y) = \Phi_M^2(g(x), y)$ holds for all $x, y \in \mathbb{N}$.

We also note that the power and the efficiency of a register machine $M$ depends on the set of instructions that are used. In [8] several sets of instructions are investigated. In particular, it is shown that there is a strongly universal register machine with 32 instructions of form $[RiP]$, $\langle Ri \rangle$, and $[RiM]$. Moreover, this machine can be effectively constructed. We rewrite the corresponding construction in terms of $[RiP]$ and $\langle RiZM \rangle$ instructions and this gives a universal register machine with 22 instructions depicted on Figure 1.

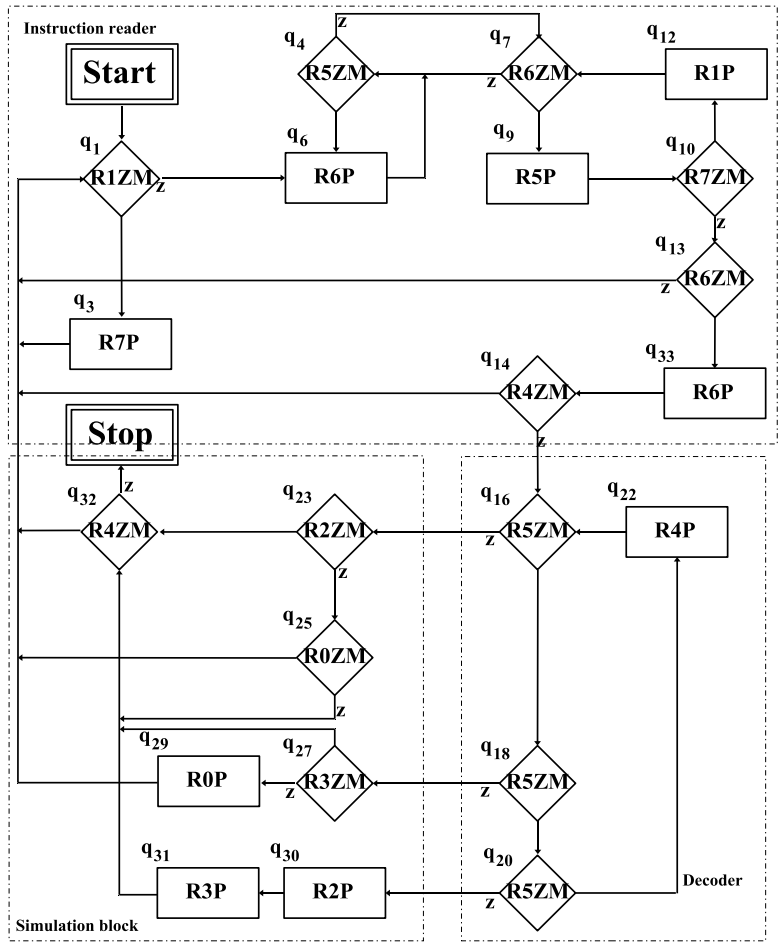

Figure 1: Flowchart of the strongly universal machine $U_{32}$ with $[RiP]$ and $\langle RiZM \rangle$ instructions.

Here is the list of rules of this machine.

$$
\begin{array}{lll}
(q_1, \langle R1ZM \rangle, q_3, q_6) & (q_3, [R7P], q_1) & (q_4, \langle R5ZM \rangle, q_6, q_7) \\
(q_6, [R6P], q_4) & (q_7, \langle R6ZM \rangle, q_9, q_4) & (q_9, [R5P], q_{10}) \\
(q_{10}, \langle R7ZM \rangle, q_{12}, q_{13}) & (q_{12}, [R1P], q_7) & (q_{13}, \langle R6ZM \rangle, q_{33}, q_1) \\
(q_{33}, [R6P], q_{14}) & (q_{14}, \langle R4ZM \rangle, q_1, q_{16}) & (q_{16}, \langle R5ZM \rangle, q_{18}, q_{23}) \\
(q_{18}, \langle R5ZM \rangle, q_{20}, q_{27}) & (q_{20}, \langle R5ZM \rangle, q_{22}, q_{30}) & (q_{22}, [R4P], q_{16}) \\
(q_{23}, \langle R2ZM \rangle, q_{32}, q_{25}) & (q_{25}, \langle R0ZM \rangle, q_1, q_{32}) & (q_{27}, \langle R3ZM \rangle, q_{32}, q_1) \\
(q_{29}, [R0P], q_1) & (q_{30}, [R2P], q_{31}) & (q_{31}, [R3P], q_{32}) \\
(q_{32}, \langle R4ZM \rangle, q_1, q_f) & &
\end{array}
$$

*2.2. Maximally parallel multiset rewriting system*

A *maximally parallel multiset rewriting system* (MPMRS) is the construct

$$\gamma = (O, \mathcal{I}, \mathcal{P}),$$

where $O$ is an alphabet, $\mathcal{I}$ is the initial multiset and $\mathcal{P}$ is a set of multiset rewriting rules (productions) of form $u \to v$, $u \in O^+$, $v \in O^*$. We say that a rule $r \in \mathcal{P}$, $r : u \to v$, is *applicable* to a multiset $X \in O^+$ if $X \supseteq u$. Similarly, a set of rules $r_i : u_i \to v_i$, $1 \leq i \leq n$ is said to be applicable to $X$ if $\sum_{1 \leq i \leq n} u_i \subseteq X$. If a rule $r \in \mathcal{P}$ is applicable to a multiset $X \in O^+$ then we define the *application* of $r$ to $X$ which produces a new multiset $Y \in O^*$; this is denoted by $X \xrightarrow{r} Y$. More exactly,

$$X \xrightarrow{r} Y \iff Y = X - u + v, \text{ where } r \text{ is } u \to v.$$

A *maximally parallel transition*, written as $X \Rightarrow Y$, is performed if there are multisets $X_1, \ldots, X_{n-1}$, $n > 0$ such that $X \xrightarrow{r_1} X_1 \xrightarrow{r_2} X_2 \xrightarrow{r_3} \cdots \xrightarrow{r_{n-1}} X_{n-1} \xrightarrow{r_n} Y$ and $r_1, \ldots, r_n \in \mathcal{P}$ is a non-deterministically chosen maximally parallel multiset of rules applicable to $X$, *i.e.*, there is no $r \in \mathcal{P}$ such that $r, r_1, \ldots, r_n$ is applicable to $X$. In a more formal way,

- $\sum_{i=1}^{n} u_i \subseteq X$,

- $u \not\subseteq X - \sum_{i=1}^{n} u_i$, for any rule $u \to v \in \mathcal{P}$,

- $Y = (X - \sum_{i=1}^{n} u_i) + \sum_{i=1}^{n} v_i$.

The first condition indicates that these rules are applicable in parallel, *i.e.*, the rules rewrite disjoint submultisets of $X$. The second condition is the maximality: no other rule is applicable in parallel with them.

By $\Rightarrow^*$ we denote the reflexive and transitive closure of $\Rightarrow$.

We may perform a maximally parallel transition on a multiset $X \in O^*$ using the following non-deterministic algorithm.

*Algorithm 1.*

> Initially the list $\mathcal{L} \in \mathcal{P}^*$ is empty and the multiset $X'$ over $O \cup \bar{O}$ is equal to $X$. Consider also the unmarking function $\mathfrak{u}(\bar{a}) = a$, $\mathfrak{u}(a) = a$, $a \in O$ and its extension to multisets. Consider $\bar{\mathcal{P}} = \{u \to \bar{v} \mid u \to v \in \mathcal{P}\}$ and suppose that rules from $\bar{\mathcal{P}}$ are ordered with respect to a total order $<$.
>
> 1. Take (non-deterministically) a rule $r \in \bar{\mathcal{P}}$ applicable to $X'$ ($r : u \to \bar{v}$).
> 2. If the previous step was successful then update $X'$ and $\mathcal{L}$: $X' = X' - u + \bar{v}$ and $\mathcal{L} = \mathcal{L}, r$. After that go to step 1.
> 3. If no rule in $\bar{\mathcal{P}}$ is applicable to $X'$, inspect $\mathcal{L} = r_1, \ldots, r_l$, $l > 0$. If the rules appear in $\mathcal{L}$ according to the order $<$, i.e. $r_i \not< r_j$ when $i > j$, then return $Y = \mathfrak{u}(X')$.
> 4. Otherwise, fail.

We recall that the result of the application of a non-deterministic algorithm is defined for non-failure cases only. The order relation is used only to avoid permutations of rules, because this leads to the same evolution.

It is clear that $X \Rightarrow Y$ is a maximally parallel transition. We also define the set

$$NEXT(X) = \{Y \mid \text{Algorithm 1 returns } Y \text{ on input } X\}.$$

It should be clear that Algorithm 1 is not the only way to compute the set $NEXT(X)$, in particular deterministic algorithms leading to the same result can be given.

We define the set of *configurations* $SF(\gamma)$ as

$$SF(\gamma) = \{w \mid \mathcal{I} \Rightarrow^* w\}.$$

We introduce the following additional notions. The *size* of a rule $u \to v$ is $|uv|$, i.e., the size of the multiset $uv$. Let $\bar{O} = \{\bar{a} \mid a \in O\}$ be the set of marked symbols from $O$ and $\mathfrak{u}$ be the unmarking morphism defined as in Algorithm 1. We say that a multiset $X$ over $O \cup \bar{O}$ is *stable* with respect to a set of rules $R$ if no rule can be applied to it: $u \not\subseteq X$ for all $u \to v \in R$.

The result of the computation of $\gamma$ is defined as

$$L(\gamma) = \{w \mid \mathcal{I} \Rightarrow^* w \text{ and } w \text{ is stable with respect to } \mathcal{P}\}.$$

*2.3. State configurations*

Now we distinguish an alphabet $R \subseteq O$ that we call the alphabet of *registers* or the *data* alphabet. A *state* configuration is the projection of a configuration over $O \setminus R$ (hence the symbols over the registers alphabet are not included in the state configuration). A state configuration $B$ is *reachable* in one step from the state configuration $A$ if there are multisets $R', R''$ over $R$ such that there exists a maximally parallel transition $AR' \Rightarrow BR''$. We will denote this by $A \Rrightarrow B$. We remark that there might be several configurations reachable in one step from a particular configuration $A$. In the general case, the number of possible state configurations is not bounded, however we would like to consider MPMRS with a finite number of state configurations.

A *finite state maximally parallel multiset rewriting system* FsMPMRS is a tuple

$$\gamma = (O, R, R_t, \mathcal{I}, \mathcal{P}),$$

where $R \subsetneq O$ and $R_t \subseteq R$ and are the alphabet of registers and the terminal alphabet of registers respectively and where $\gamma' = (O, \mathcal{I}, \mathcal{P})$ is an MPMRS which has a finite number of state configurations, *i.e.*, the projection of $SF(\gamma')$ over $O \setminus R$ is finite. Moreover, we require that for any rule $r \in \mathcal{P}$, $r : u \to v$, $u$ must contain at least one symbol from $O \setminus R$.

The result of the computation of $\gamma$ is the projection of $L(\gamma')$ over $R_t$:

$$L(\gamma) = \pi_{R_t}(L(\gamma')).$$

We would like to note that FsMPMRS uses the paradigm of imperative programming where the finite control (the program) is separated from the data, which differentiates it from MPMRS or ARMS having an unbounded control. In our case the state configurations correspond to the program and the projection of a configuration over $R$ corresponds to the data. Moreover, the restriction on the rules implies that the data cannot evolve by itself. This is similar to many computational models, for example, in case of register machines there is a strict separation between states and registers, and the registers cannot evolve by themselves.

We say that a rule $u \to v$ is a *pure state* rule if $u$ contains no symbols from $R$, otherwise we call it a *register-dependent* rule.

The set of state configurations of an FsMPMRS may be computed iteratively as follows (where $R^\infty$ denotes the infinite multiset over $R$):

1. $C_0 = \{\mathcal{I}\}$.
2. $C_{i+1} = C_i \cup \{\pi_{O \setminus R}(Y) \mid Y \in NEXT(X + R^\infty) \text{ for all } X \in C_i\}$.

We remark that $NEXT(X + R^\infty)$ is finite because there are no rules that involve only registers. In fact, $NEXT(X + R^\infty)$ may be obtained by computing $NEXT(X)$ for a system where register symbols in the rules are ignored. Moreover, we notice that actual infinity of register symbols is not needed, since the maximal number of them that can be consumed does not exceed the maximal number of register symbols that appear in the left hand side of a rule, multiplied by size of $X$. Therefore, when speaking about applicability of rules without worrying about the register symbols, we will write $X + R^\infty$, in the context of these observations.

*2.3.1. Relation to antiport P systems*

We recall here the definition of a symport/antiport P system.

**Definition 1.** A P system with symport/antiport of degree $n$ is a construct

$$\Pi = (O, \mu, w_1, \ldots, w_n, E, R_1, \ldots, R_n, i_0),$$

where:

1. $O$ is a finite alphabet of symbols called objects,
2. $\mu$ is a membrane structure consisting of $n$ membranes that are labeled in a one-to-one manner by $1, 2, \ldots, n$. Usually, $\mu$ is given as a set of nested parentheses.
3. $w_i \in O^*$, for each $1 \leq i \leq n$ is a finite multiset of objects associated with the region $i$ (delimited by membrane $i$),
4. $E \subseteq O$ is the set of objects that appear in the environment in infinite numbers of copies,
5. $R_i$, for each $1 \leq i \leq n$, is a finite set of symport/antiport rules associated with the region $i$ and which have the following form $(x, in), (y, out), (y, out; x, in)$, where $x, y \in O^*$,

6. $i_0$ is the label of a membrane of $\mu$ that identifies the corresponding output region.

A symport/antiport P system is defined as a computational device consisting of a set of $n$ hierarchically nested membranes that identify $n$ distinct regions (the membrane structure $\mu$), where to each region $i$ there are assigned a multiset of objects $w_i$ and a finite set of symport/antiport rules $\mathcal{R}_i$, $1 \leq i \leq n$. A symport rule $(x, in) \in \mathcal{R}_i$ permits to move $x$ into region $i$ from the immediately outer region. Notice that rules of the form $(x, in)$, where $x \in E^*$ are forbidden in the skin (the outermost) membrane. A symport rule $(x, out) \in R_i$ permits to move the multiset $x$ from region $i$ to the outer region. An antiport rule $(y, out; x, in)$ exchanges two multisets $y$ and $x$, which are situated in region $i$ and the outer region of $i$ respectively.

A computation in a symport/antiport P system is obtained by applying the rules in a non-deterministic maximally parallel manner, i.e. all rules that can be applied together should be applied. The computation is restricted to moving objects through membranes, since symport/antiport rules do not allow the system to modify the objects placed inside the regions. Initially, each region $i$ contains the corresponding finite multiset $w_i$; whereas the environment contains only objects from $E$ that appear in infinitely many copies.

A computation is successful if starting from the initial configuration it reaches a configuration where no rule can be applied. The result of a successful computation is the natural number that is obtained by counting the objects that are present in region $i_0$. Given a P system $\Pi$, the set of natural numbers computed in this way by $\Pi$ is denoted by $N(\Pi)$. It is also possible to count only symbols over a specified alphabet $T$ as a result. The corresponding way of taking the result will be denotes as $N_T(\Pi)$. We shall use in what follows antiport P systems, *i.e.*, the symport rules are not considered.

An antiport P system with one membrane $\Pi = (O, [\,]_1, w_1, E, R_1, 1)$ can be represented as a FsMPMRS system $\gamma = (O, E, O \setminus E, w_1, \mathcal{P})$, where $\mathcal{P} = \{u \to v \mid (u, out; v, in) \in R_1\}$. Clearly, the converse also holds: a FsMPMRS system $\gamma = (O, R, R_t, \mathcal{I}, \mathcal{P})$ can be represented as a an antiport P system with one membrane $\Pi = (O, [\,]_1, \mathcal{I}, R, R_1, 1)$, where $R_1 = \{(u, out; v, in) \mid u \to v \in \mathcal{P}\}$. The projection over $R_t$ can be obtained by taking $N_{R_t}(\Pi)$.

*2.3.2. Graphical notation*

We introduce a graphical notation for FsMPMRS. We represent a state configuration by a filled square. We also suppose that pure state rules precede register-dependent rules. Now, in order to represent the relations between state configurations we will depict the relation $\Rrightarrow$ by graphically representing the functioning of Algorithm 1. We take a state configuration $X$ and apply the algorithm for the multiset $X + R^\infty$. It is clear that for any positive run of the algorithm that returns the multiset $Y$ with $\mathcal{L} = r_1, \ldots, r_n$ the equation $X + R^\infty \xrightarrow{r_1} X_1 \xrightarrow{r_2} X_2 \xrightarrow{r_3} \cdots \xrightarrow{r_n} X_n$ holds, where $\mathfrak{u}(X_n) = Y$. More precisely, multisets $X_i$ over $O \cup \bar{O}$, $1 \leq i \leq n$ are obtained in the second step of the algorithm. We take the projection over $O \cup \bar{O} \setminus (R \cup \bar{R})$ of each of these intermediate multisets and represent it by a circle. We also draw an arrow labeled by $r_i$ between circles corresponding to $\pi_{O \cup \bar{O} \setminus (R \cup \bar{R})}(X_{i-1})$ and $\pi_{O \cup \bar{O} \setminus (R \cup \bar{R})}(X_i)$. We also depict beside the arrow all symbols over $R$ from the right-hand (resp. left-hand) side of $r_i$ preceding them with the + (resp. -) sign. Finally, we attach by a line the circle corresponding to a configuration $Z$ to the square $\mathfrak{u}(Z)$. If all circles attached to a square represent multisets over $O \cup \bar{O}$ that are not stable with respect to pure state rules, such square is not filled. Several circles attached to the same square and not having outgoing arrows can be combined to a single circle.

The final diagram is obtained by repeating the above construction for all possible runs of the Algorithm 1 and for all state configurations. We recall that there is a finite number of state configurations and a finite number of possible runs of the Algorithm 1, hence the above process will stop at some moment.

**Example 1.** Consider the following system $\gamma = (\{A, B, C, D\}, \{E, F\}, \{F\}, \{AABEE\}, \mathcal{P})$, where $\mathcal{P}$ contains the following rules:

$$r_1 : AB \to C$$
$$r_2 : AE \to D$$
$$r_3 : DC \to AABF$$

Clearly, the system $\gamma$ is an FsMPMRS that computes the multiset $\{FF\}$. Indeed, there are three state configurations $AAB$, $AC$ and $CD$ and there are no rules involving only $E$ or $F$ in the left-hand side. In a graphical way this system is represented as follows:

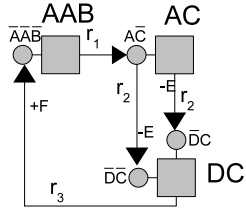

The graphical notation described above not only describes the functioning of Algorithm 1, but also gives a tool which may be used for the description of the evolution of the system. Consider an arbitrary transition $X \Rightarrow Y$ corresponding to a parallel application of rules $r_1, \cdots, r_n$. If $X \xrightarrow{r_1} X_1 \xrightarrow{r_2} \cdots \xrightarrow{r_n} X_n = Y$, this can be graphically followed as a path from a square corresponding to $\pi_{O \setminus R}(X)$, going through circles corresponding to $\pi_{O \cup \bar{O} \setminus (R \cup \bar{R})}(X_i)$, $1 \leq i \leq n$, and the last one is attached to a square corresponding to $\pi_{O \setminus R}(Y)$. The maximality of parallelism should translate in the following way: no rules corresponding to arrows from the circle corresponding to $\pi_{O \cup \bar{O} \setminus (R \cup \bar{R})}(X_n)$ should be applicable to $X_n$.

Therefore, applying any maximally parallel transition to a configuration $X$ means to start from the square $\pi_{O \setminus R}(X)$ and follow arrows to circles as long as possible, keeping track of symbols from $R$; when it is no longer possible, consider the square to which the last circle is attached.

For the previous example, in order to compute the first step using the diagram we proceed as follows. The initial configuration $AABEE$ corresponds to the state configuration $AAB$. Now rule $r_1$ can be applied bringing us to the circle attached to the square $AC$. At this moment our intermediate configuration is $\bar{A}A\bar{B}EE$. Since $r_2$ is applicable we can continue to the circle attached to the square $DC$ and take into account that one $E$ is removed. At this moment our intermediate configuration is $\bar{A}\bar{A}\bar{B}\bar{E}E$. Finally, since there are no more outgoing arrows, we stop at the square $DC$ and our configuration is $DCE$.

Taking into account the above description, we can simplify the diagram from the example above by the following observation. For any configuration $X$ having $\pi_{O \setminus R}(X) = AAB$, if rule $r_2$ is not applicable from the circle $A\bar{C}$, then it will not be applicable from the square $AC$ because otherwise it would have been applied in the previous step. Hence, it may be eliminated:

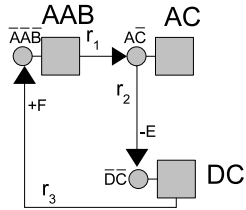

This can be formalized as follows.

**Proposition 1.** *If the following conditions hold:*

- *A is a state configuration and B and C are two state configurations reachable in one step from A;*
- *there is a path (a sequence of rules) obtained by Algorithm 1 $A + R^\infty \to^* B' \xrightarrow{r_1} \cdots \xrightarrow{r_n} C'$ such that $\pi_{O\setminus R}(\mathfrak{u}(B')) = B$ and $\pi_{O\setminus R}(\mathfrak{u}(C')) = C$;*
- *there is no state configuration D other that those on the mentioned path, such that B is reachable in one step from D;*

*then the path labeled by $r_1, \ldots, r_n$ leading from B to C may never be involved in a computation and it may be eliminated from the diagram.*

We remark that the introduced graphical notation is different from the Molecular Interactions Maps [7] or Kitano [6] notations, used for specifications in systems biology, because it permits to represent the computational flow of the system.

## 3. The Basic Simulation Technique

In this section we concentrate on a simple simulation of register machines by FsMPMRS. This simulation is done as follows. We represent the current configuration of a register machine $M$ by a multiset (initially $\mathcal{I}$). In particular, the contents of a register $i$ is represented by the number of symbols $R_i$ which are present. The simulation of any incrementing or decrementing instruction of $M$ is done by an appropriate set of rules.

In order to construct an MPMRS with a small number of rules we shall follow ideas presented in [5]. We take them as a starting point and after that we consider different minimization strategies that will decrease the number of used rules.

The system from [5] is based on a simulation of a special universal register machine $U_{32}$ having 32 instructions taken from [8]. This construction may be rewritten in terms of $[RiP]$ and $\langle RiZM \rangle$ instructions, which gives 22 rules (9 incrementing instructions and 13 decrementing), see Figure 1.

The basic simulation strategy consists in a simulation of rules of an arbitrary register machine by multiset rewriting rules using the smallest number of the latter ones. Any incrementing rule $(q, [RiP], q_1)$ of register machine can be directly simulated by the rule

$$q \to R_i q_1, \tag{1}$$

This corresponds to the following flowchart:

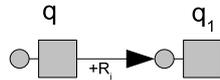

Any decrementing rule $(q, \langle Ri_q ZM \rangle, q_1, q_2)$ can be simulated using five rules:

$$\begin{aligned} q &\to q' C_q, \\ q' &\to q'', & C_q R_{i_q} &\to C'_q, \\ q'' C_q &\to q_1, & q'' C'_q &\to q_2 \end{aligned} \tag{2}$$

This corresponds to the following flowchart:

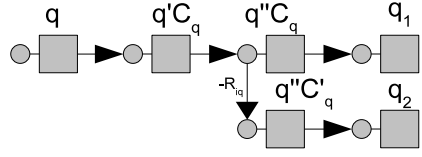

This simulation is done as follows. Symbol $q$ introduces symbols $q'$ and $C_q$ (the last one is called the *checker* for the state $q$).

After that symbol $C_q$ tries to decrease register $i_q$ and if it succeeds then it becomes $C'_q$. Now, depending on this information symbol $q''$, which replaced $q'$, will choose the corresponding new state.

The choice between configurations $q''C_q$ and $q''C'_q$ depends on the presence of symbol $R_{i_q}$, i.e., if register $i_q$ is zero.

Applied to $U_{32}$ this translation gives an FsMPMRS with 73 rules. We remark that these rules are of size at most 3. In the following sections we show different techniques which reduce the number of rules for the price of increasing their size.

## 4. Basic minimization strategies

In this section we present two basic minimization strategies. One of them is based on structural improvements and the other one is based on encodings. We present them in a general form and after that we show how they apply to the system that simulates $U_{32}$.

*4.1. State Elimination*

This minimization strategy performs an elimination of linear fragments in the flow-chart (by performing a kind of speed-up). Suppose that there are following two pure state rules, $r_1 = (q_1 \to q_2)$ and $r_2 = (q_2 \to q_3 R_i)$. This corresponds to the flowchart in the picture.

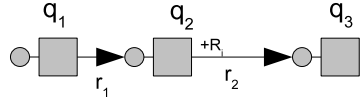

We observe that rules $r_1$ and $r_2$ may be combined and state $q_2$ may be eliminated by introducing a new rule $r = (q_1 \to q_3 R_i)$. In a similar way, any linear chain of pure state rules may be collapsed to a single rule (the size may increase for each additional rule). We shall further refer to this technique as *intermediate state elimination*.

For $U_{32}$ we observe that using intermediate state elimination technique we can reduce (2) to following rules (we also renamed $q'$ to $q$ and assume that the initial state is encoded $q_o C_{q_0}$):

$$q \to q', \qquad\qquad C_q R_{i_q} \to C'_q, \qquad (3)$$
$$q'C_q \to q_1 C_{q_1}, \qquad\qquad q'C'_q \to q_2 C_{q_2}$$

Graphically this can be represented as follows:

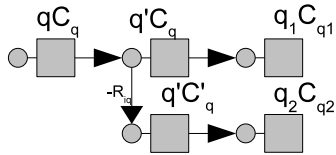

We observe that compared to the previous picture of the decrement simulation the state $q$ was eliminated and now the simulation starts with the state $qC_q$ (previously $q'C_q$).

Moreover, we observe that for $U_{32}$ in most cases a decrementing instruction $(q, \langle RiZM \rangle, q_1, q_2)$ is followed by an incrementing instruction $(q_1, [Rk_1P], q_3)$ or $(q_2, [Rk_2P], q_4)$. Hence, one can simulate the incrementing instruction during the simulation of the previous decrementing instruction (by eliminating the unneeded state in between). For example, last two rules from (3) become

$$q'C_q \to q_3 R_{k_1} C_{q_3}, \qquad\qquad q'C'_q \to q_4 R_{k_2} C_{q_4}. \qquad(4)$$

Of course, this increases the size of rules up to 5.

### 4.2. Gluing rules

In this section we shall consider techniques that will minimize the number of rules by performing more transitions between the configurations by fewer rules. Informally, transitions $c_1 \overset{r1}{\to} c_2$ and $d_1 \overset{r2}{\to} d_2$ can be performed by the same rule $X \to Y$ if they are represented in a suitable way: $c_1 = cX$, $c_2 = cY$, $d_1 = dX$, $d_2 = dY$. In this case, we say that $r_1$ and $r_2$ may be *glued*. The following picture illustrates this:

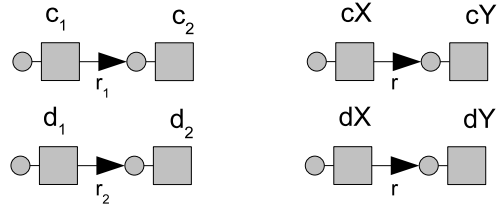

In a more formal way one must find a suitable encoding of state configurations such that:

- No state configuration is a submultiset of another state configuration.

- There should be at least 2 transitions that may be glued.

We would like to remark that it is only possible to glue transitions that increment registers equivalently, in particular, transitions that do not increment any register.

In what follows we apply the idea of gluing rules to the FsMPMRS system obtained by the basic simulation technique.

#### 4.2.1. Phases

Consider now the rules (3). If we represent the state $q$ by $qS$ and the state $q'$ by $qS'$ then the first rule from (3) may be glued for all states $q$, i.e., instead of $|Q|$ rules $q \to q'$ we obtain one rule $S \to S'$. We call the symbol $S$ the *phase*, hence there will be two phases $S$ and $S'$. The rules from (3) are replaced by:

$$C_q R_{i_q} \to C'_q, \qquad(5)$$
$$qS'C_q \to q_1 C_{q_1} S, \qquad\qquad qS'C'_q \to q_2 C_{q_2} S$$

Graphically this is represented as follows (where the double-headed arrow represents the rule $S \to S'$ common for all simulation blocks):

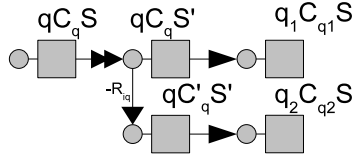

### 4.2.2. Independent Checkers

Another minimization idea comes from the observation that the information encoded in the checker $C_q$ from (5) is redundant. If we take the set of first rules from (5) for all $q \in Q$, we observe that it is possible to glue rules that decrement the same register in the following way. We encode the multiset $qC_q$, respectively $qC'_q$, by symbols $qC_{i_q}$, respectively $qC'_{i_q}$, where $i_q$ is the number of the register decreased by the instruction $q$ of $M$. Now it is possible to eliminate the first rule from (5) by introducing rules $C_i R_i \to C'_i$, $1 \le i \le |R|$. By convention, we will say that a state $q$ of machine $M$ is encoded by symbols $qC_{i_q}S$ and we will say that $C_{i_q}$ is the checker for the state $q$. This transforms (5) into the following:

$$qS'C_{i_q} \to q_1 C_{i_{q_1}} S, \qquad\qquad qS'C'_{i_q} \to q_2 C_{i_{q_2}} S, \qquad\qquad (6)$$

where $C_{i_{q_1}}$ and $C_{i_{q_2}}$ represent checkers for states $q_1$ and $q_2$. Of course this introduces $|R|$ new rules, but finally we gain more because of the elimination of one rule for each $q \in Q$ from (5).

Graphically this is represented at follows:

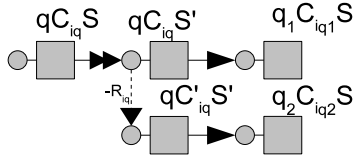

### 4.3. Remarks

The improvements to the $U_{32}$ simulation presented above were implemented in [5], but they were classified in a different way. The cited article considers the relation between the size of rules and their number. In the table below we collect the results obtained by using the techniques above, as well as results from paper [5] which uses ideas similar to those used in the current and the next section. To highlight the results from [5] we underline them:

| Size | Number of rules |
|------|-----------------|
| 3    | <u>73</u>       |
| 5    | 56              |
| 6    | 47              |
| 7    | <u>43</u>       |
| 11   | <u>30</u>       |
| 20   | 23              |

## 5. Further Minimization of $U_{32}$ Simulation

In this section we show how to minimize the simulation of $U_{32}$. We start with the simulation using rules (6) and we do structural improvements based on some observations on the functioning of the system. After that we show how to glue most of the remaining rules by showing a suitable encoding of state configurations.

*5.1. Structural improvements*

The structural improvements presented in this section are in some sense a generalization of the intermediate state elimination technique.

*5.1.1. Reducing decoder block*

The first important improvement may be done by considering the decoder part of the machine from [8] (see the flowchart in Figure 1). In fact, this block does a division of register 5 by three. The result of this division is stored in register 4 and according to the value of the remainder states $q_{23}$, $q_{27}$ and $q_{30}$ are chosen respectively. This behavior may be simulated by 5 rules which try to decrease register 5 by 3 and make the choice depending on the result of this subtraction. The state $q_{16}$ is now encoded by $q_{16}C_5C_5C_5S$.

$$C_5R_5 \to C_5',$$
$$q_{16}C_5C_5C_5 \to q_{23}C_2S, \qquad q_{16}C_5C_5C_5' \to q_{27}C_3S \qquad (7)$$
$$q_{16}C_5C_5'C_5' \to q_{32}C_4R_3R_2S, \qquad q_{16}C_5'C_5'C_5' \to q_{16}C_5C_5C_5R_4S$$

We note that we combined 2 addition instructions in the third branch using the elimination of intermediate states ($q_{30}$ and $q_{31}$) by the mechanism discussed above. The subtraction by 3 is done using the maximal parallelism which permits to apply the rule $C_5R_5 \to C_5'$ three times at least 3 copies of $R_5$ are available. In [5] the idea of checking several registers at the same time is developed in more details, however here we will use another structural idea which is more efficient.

We integrally present the obtained flowchart in Figure 2. We use following conventions. The double-headed arrow represents the rule $S \to S'$ that changes the phase. Rules that decrement registers ($C_iR_i \to C_i'$) are represented by arrows starting with a perpendicular bar and labeled by $D0$–$D7$ enclosed in circle. We also do not depict in this case the corresponding decrementing register. Rules that increment registers are depicted by arrows with a dashed line and the incremented register(s) are depicted beside the line. These rules are labeled by a letter enclosed in a diamond. All other rules (which do not increment/decrement registers) are labeled by a number enclosed in a square.

*5.1.2. State Elimination for Decrementing Rules*

If we look at the flowchart in Figure 2 we observe that rules $D0$–$D3$ and $D7$ are used only once. This gives the possibility to combine these rules with rules that follow them using the state elimination technique. For example, the transformation done by rule 9 may be done directly in rule $D2$. However, in this case we reach the configuration $q_{32}C_4S'$ instead of $q_{32}C_4S$ because the rule $S \to S'$ is performed independently. We may solve this issue by introducing 3 phases instead of 2. In this case, phase 2 (marked by $S'$) will be treated analogously to phase 1 (marked by $S$) and the move to the next state will be done in phase 3 (marked by $S''$). Moreover, the phase change may still be done by one rule. For this it is enough to replace $S$ by $XXX$, $S'$ by $XXT$ and $S''$ by $XTT$ and the rules $S \to S'$ and $S' \to S''$ by the rule $XX \to XT$. These changes permit to save 3 rules because rule $D1$ is a special case and it cannot be combined with rule $h$. However, we can include the increment of register 7 in rule $D1$, in this case rule $h$ becomes a non-incrementing rule and it will be labeled by $1a$.

The new flowchart is shown in Figure 3. We still depict phases by symbol $S$ with primes because of the lack of the space. However, it is clear that the above substitution shall be done.

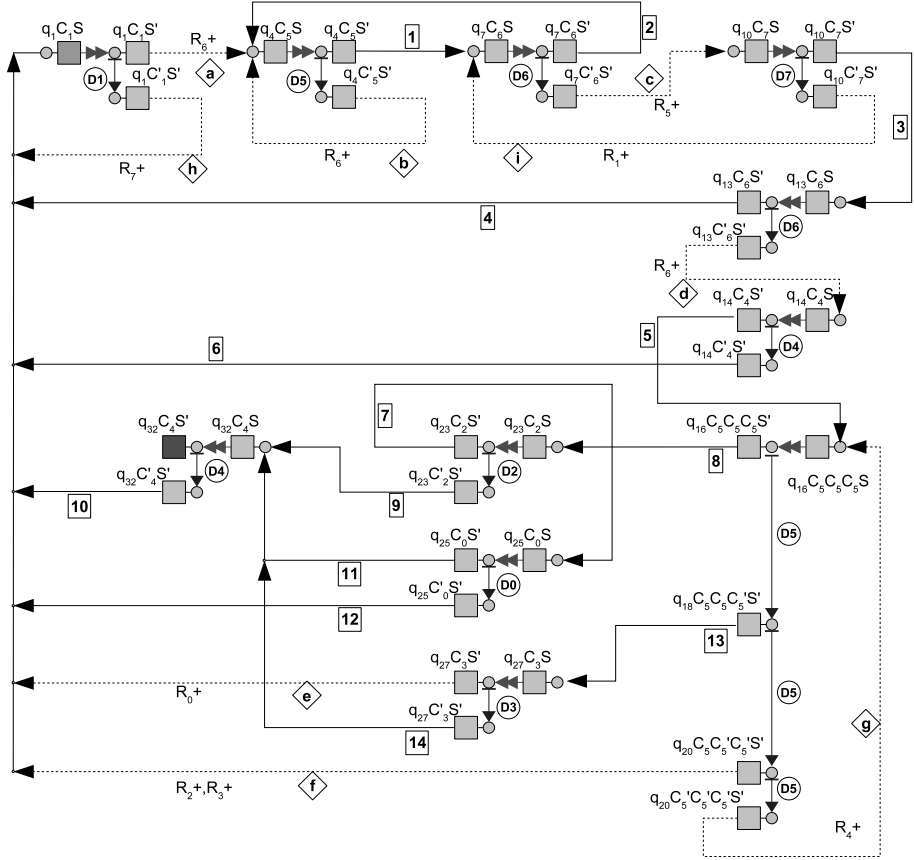

Figure 2: Multiset rewriting flowchart of $U_{32}$ with improved decoder block

*5.2. Encoding optimization*

In this section we show how using the gluing minimization strategy the number of rules may be substantially decreased.

From Figure 3 we can see that transitions $a$, $b$ and $d$ may be potentially glued together as well as transitions labeled by numbers. All other rules are not eligible for gluing. Consider the part of the flowchart that involves transitions labeled by numbers and all corresponding phases. Figure 4 depicts this. We denote rules that apply in parallel by drawing corresponding arrows beside each other and we do not show the $S'$ phase (because in our case it differs from the $S''$ phase only by the phase symbol).

We remark that all rules labeled by numbers change the phase from $S''$ to $S$. We call such rules *phase changing* rules, in contrast to *non-phase* rules like $S \to S'$ (and $S' \to S''$) or $C_i R_i \to C'_i$. It is clear that non-phase rules may performed in parallel to each other or to a phase changing rule.

We found two strategies that permit to create suitable encodings in order to glue rules:

- End of phase discrimination.

- Maximally parallel unification.

The first strategy consists in assigning one phase changing rule per each arrow that enters a node with the maximal number of entering arrows. After that all other arrows are identified with one of the obtained arrows and the corresponding encoding is deduced.

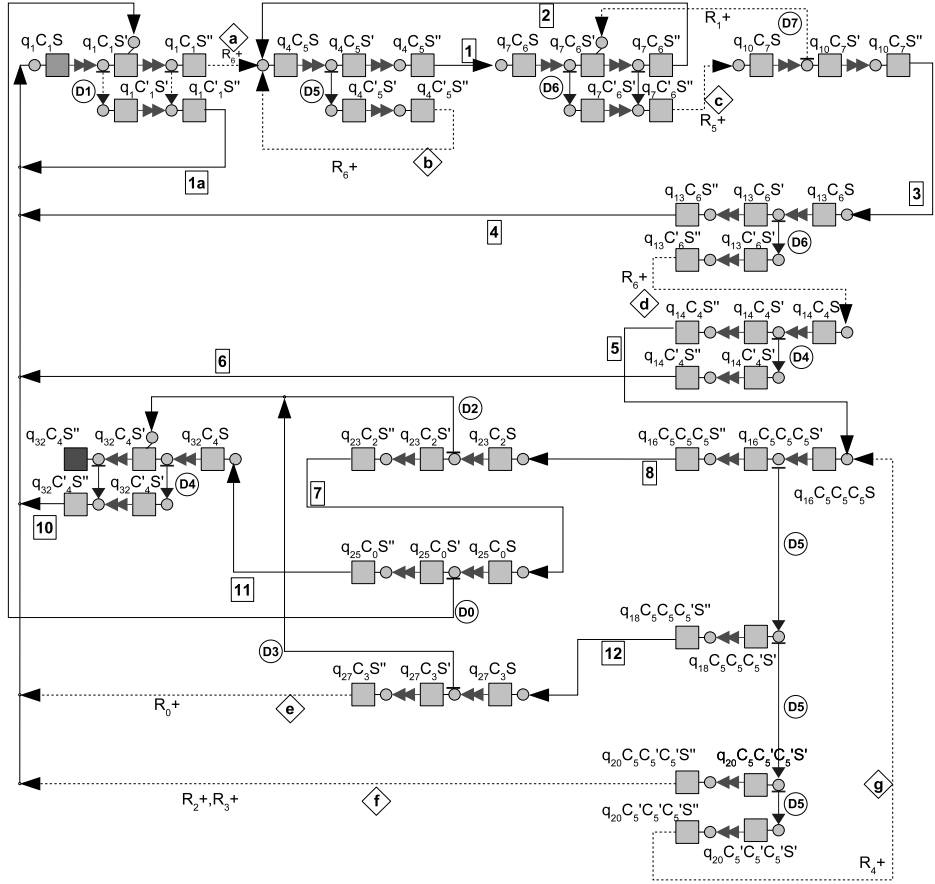

Figure 3: Multiset rewriting flowchart of $U_{32}$ with 3 phases

The second technique permits to reduce the number of phase changing rules in the maximal node (and potentially in some other distant nodes). This technique is based on the fact that any number of phase changing rules that enter a node $U$ may be reduced to only two rules, one phase changing and one non-phase rule. More precisely, we number rules from 0 to $k$, where $k$ is the number of entering phase changing rules and denote by $U_j$, $0 \leq j \leq k$ corresponding nodes. After that we introduce new special symbols $A^j B^{k-j}$ in the encoding of $U_j$. We also consider that $B^k$ is a part of the encoding of $U$. Such an encoding permits to make the transition from any of $U_j$, $0 \leq j \leq k$ to $U$ by applying an appropriate phase changing rule and several parallel applications of the non-phase rule $A \to B$.

However, this technique has some limitations. In fact, it is applicable only if rules that are glued are preceded by rules that introduce the difference between encodings which is subject to the non-phase rule (the symbols $A$ and $B$). In the case of our system, inherited from the basic simulation of $U_{32}$, this must be a rule that decrements a register. This is due to the fact that the non-phase rule may be applied in parallel to any other rule, so corresponding symbols must be introduced just before making the last step, namely during phase 2. If these symbols are introduced during phase 1, then in the case if corresponding computational branch is not chosen, for example if register is zero, then a wrong encoding (corresponding to some other state) is produced.

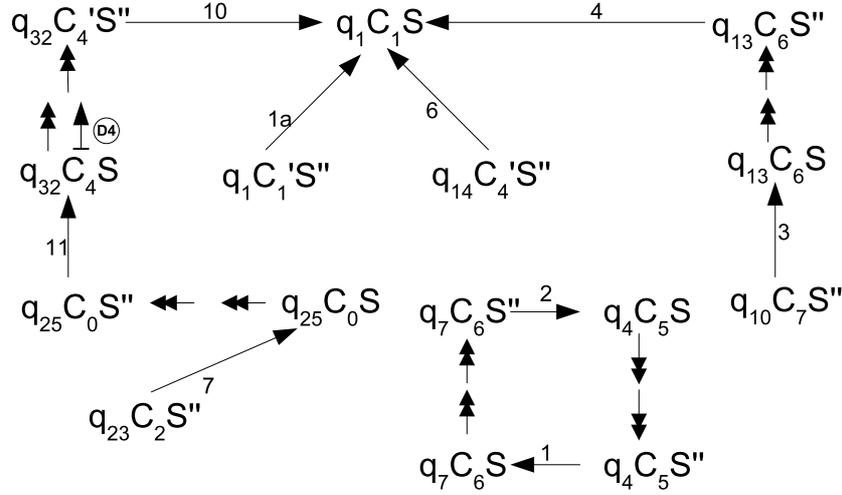

Figure 4: Part of the multiset rewriting flowchart of $U_{32}$ showing only rules labeled by numbers.

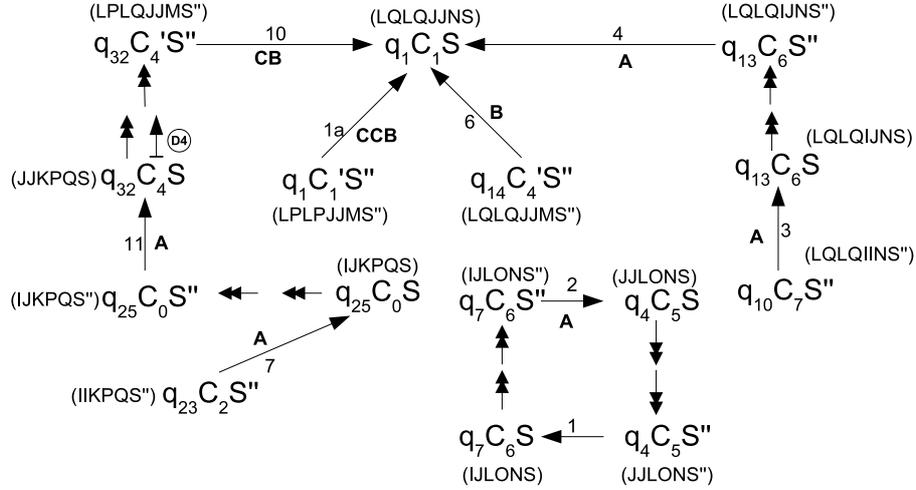

Figure 5: Part of the multiset rewriting flowchart of $U_{32}$ showing only glued rules and the corresponding encoding.

After applying all considerations above we found an encoding of the part of the flowchart represented in Figure 4 that permits to perform 9 rules (labeled by numbers) by only 3 new rules. This encoding is shown in Figure 5 and it is based on 2 phase changing rules $A = (IS'' \to JS)$ and $B = (JJMS'' \to JJNS)$ and on one non-phase rule $C = (LP \to LQ)$.

It was done as follows. The node $q_1 C_1 S$ has the maximal number of entering arrows, hence we try to keep only the 4 rules that enter this node (10, 1a, 6 and 4) and express all other rules in their terms. However, before doing this it is better to apply the maximally parallel unification of the above 4 rules. We remark that rules 10, 1a and 6 exit from states reached by a decrement rule, while the rule 4 exits a state that is not reached by a decrement rule. Hence, we need two phase changing rules to perform rules 6 and 4. We denote them by $B$ and $A$ respectively. Rules 1a and 10 are performed by the combination $CB$ and $CCB$ respectively. We recall that rule $C$ is performed in parallel to any other rule.

Figure 6 shows the flowchart obtained after applying all ideas mentioned before. Arrows ending with a diamond correspond to rule $C$, while arrows ending with a square correspond to rule $A$. The sparse arrow corresponds to rule $B$.

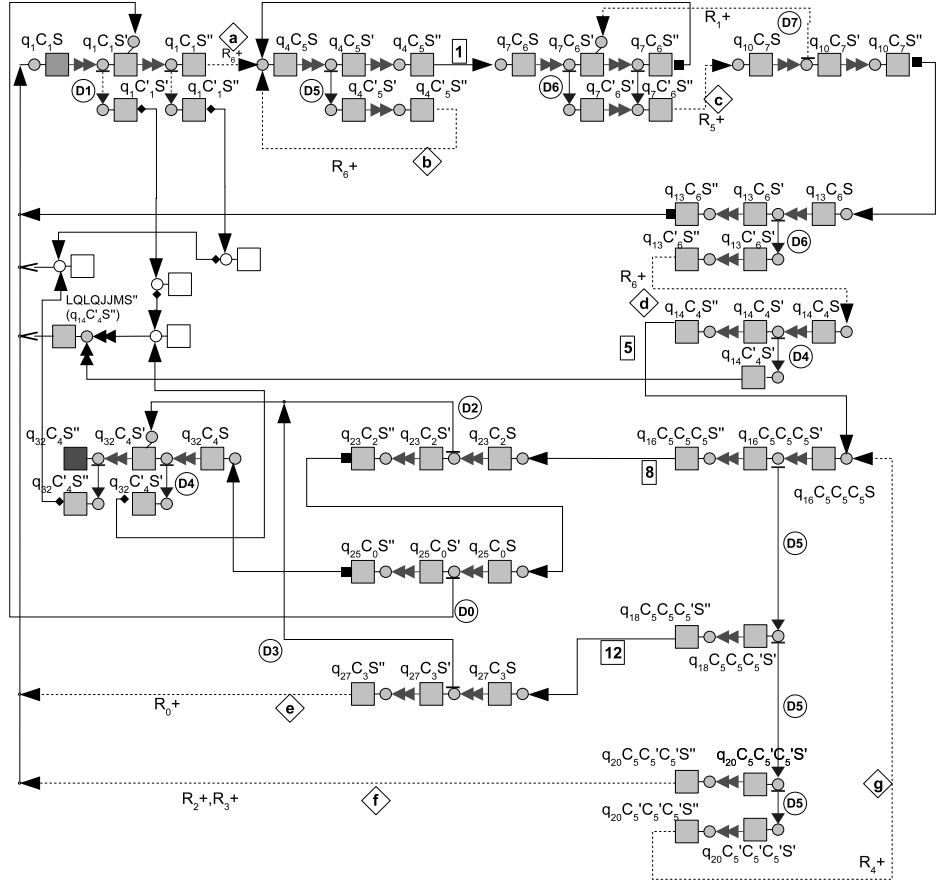

Figure 6: Multiset rewriting flowchart of $U_{32}$ with glued rules.

## 6. Formal Description of the System

In this section we give the formal description of the obtained system.

We constructed the system $\gamma = (O, R, \{R_1\}, \mathcal{I}, \mathcal{P})$, where

$$\begin{aligned}
O &= R \cup \{C_3, C'_5, C'_6\} \cup \{q_{16}, q_{27}\} \cup \{T, I, J, K, L, M, N, O, P, Q, T, X\}, \\
R &= \{R_i \mid 0 \leq i \leq 7\}, \\
\mathcal{I} &= LQLQJJNXXXR_0^{i_0} \cdots R_7^{i_7}.
\end{aligned}$$

Here $i_0, \ldots, i_7$ is the contents of registers 0 to 7 of $U_{32}$ and $LQLQJJNXXX$ is the encoding of the initial state $q_1C_1S$. The set of rules $\mathcal{P}$ is the following:

| No | Rule | No | Rule |
|---|---|---|---|
| phase | $XX \to XT$ | | |
| D0 | $IJKPQR_0 \to LQLQJJM$ | a | $LQLQJJNTT \to JJLOR_6XX$ |
| D1 | $LQLQJJNR_1 \to LPLPJJMR_7$ | b | $LC'_5TT \to JJLOR_6XX$ |
| D2 | $IIKPQR_2 \to JJKPQ$ | c | $OC'_6TT \to IILQLQNR_5XX$ |
| D3 | $q_{27}C_3R_3 \to JJKPQ$ | d | $QLQNC'_6TT \to JJKQQR_6XX$ |
| D4 | $JJKR_4 \to JJLLM$ | e | $q_{27}C_3TT \to LQLQJJNR_0XX$ |
| D5 | $JJOR_5 \to C'_5$ | f | $q_{16}JJOC'_5C'_5TT \to LQLQJJNR_2R_3XX$ |
| D6 | $IJLR_6 \to C'_6$ | g | $q_{16}C'_5C'_5C'_5TT \to q_{16}JJOJJOJJOXX$ |
| D7 | $IILQLQNR_7 \to IJLOR_1$ | | |
| A | $ITT \to JXX$ | 1 | $JJLOTT \to IJLOXX$ |
| B | $JJMTT \to JJNXX$ | 5 | $JJKQQTT \to q_{16}JJOJJOJJOXX$ |
| C | $LP \to LQ$ | 8 | $q_{16}JJOJJOJJOTT \to IIKPQMXX$ |
| | | 12 | $q_{16}JJOJJOC'_5TT \to q_{27}C_3XX$ |

As a corollary of the discussion from previous sections and the system above we obtain:

**Theorem 1.** *There exists a universal FsMPMRS having 23 rewriting rules.*

## 7. Conclusions

In this article we investigated maximally parallel multiset rewriting systems (MPMRS) which correspond in a direct way to antiport P systems with one membrane. We constructed a universal (Fs)MPMRS that computes any partially recursive function providing that the input is the encoding of a register machine computing the corresponding function as well as the value to be computed. Our construction uses 23 rules. This result is quite surprising, because the machine from [8] that was the starting point of our construction uses 25 computational branches. This means that some branches in [8] do the same thing and are maybe redundant. Hence the result of this paper may possibly help to decrease the number of rules for universal register machines.

We also introduced a graphical notation for FsMPMRS, which shows the functioning of the system, and presented two different minimization strategies for (Fs)MPMRS based on structural transformations (on the elimination of intermediate states) and on encodings (gluing rules) and discussed how their consequent application may decrease the number of rules in the simulation of $U_{32}$. These strategies may help to design more compact MPMRS and we leave the question open about a smaller number of rules for a universal MPMRS (or antiport P systems with one membrane). Moreover, we observed a trade-off between the number of rules and their size (also observed in [5]) and we think that a further study of the relation between the size of rules and their influence on different minimization strategies (and of course their number) is interesting.

*Acknowledgments.* The authors gratefully acknowledges support by the Science and Technology Center in Ukraine, project 4032. The first author also acknowledges the support of Academy of Finland, project 203667 and the Japan Society for the Promotion of Science and the Grant-in-Aid for Scientific Research, project 20·08364.